\documentclass[11pt]{article}

\usepackage[english]{babel}
\usepackage[a4paper,dvips,nohead,twoside]{geometry}
\usepackage{ucs}
\usepackage[utf8x]{inputenc}
\usepackage{amsmath,amssymb}
\usepackage[dvips]{graphicx}
\usepackage{psfrag}
\usepackage{xspace}
\usepackage{tensor}

\usepackage[longversion]{optional}

\usepackage[ps2pdf, bookmarks=true,
            bookmarksopen=true,colorlinks=false]{hyperref}

\hypersetup{%
  pdftitle = {Non-genericity of the Nariai solutions:
    I. Asymptotics and spatially homogeneous perturbations},
  pdfsubject = {},
  pdfauthor = {Florian Beyer},
  pdfkeywords = {}%
}

 \usepackage[thmmarks,amsmath,hyperref]{ntheorem}

 \newcommand{\Propref}[1]{Proposition~\ref{#1}}

 \theoremstyle{plain}
 \theorembodyfont{\normalfont}
 \theoremsymbol{}
 \newtheorem{Def}{Definition}

 \newtheorem{Prop}[Def]{Proposition}
 \newtheorem{Conj}[Def]{Conjecture}
 
 \theoremheaderfont{\sc}\theorembodyfont{\small\upshape}
 \theoremstyle{nonumberplain}
 \theoremseparator{:}
 \theoremsymbol{\rule{1ex}{1ex}}
 \theoremstyle{break}
 \theoremheaderfont{\normalfont\small}
 \theorembodyfont{\it\small\flushleft}
 \theoremsymbol{}
 
 \theorembodyfont{\it\tiny\flushleft}

\newcommand{\R}{\ensuremath{\mathbb R}\xspace}

\newcommand{\SO}{\ensuremath{\text{SO}(3)}\xspace}
\newcommand{\So}{\ensuremath{\mathbb S^1}\xspace}
\newcommand{\St}{\ensuremath{\mathbb S^2}\xspace}

\newcommand{\SoXSt}{\ensuremath{\mathbb S^1\times\mathbb S^2}\xspace}

\newcommand{\scrip}{\ensuremath{\mathcal J^+}\xspace}
\newcommand{\scrim}{\ensuremath{\mathcal J^-}\xspace}

\newcommand{\arctanh}{\ensuremath{\text{arctanh}}\xspace}

\newcommand{\Eqref}[1]{Eq.~\eqref{#1}}

\newcommand{\Sectionref}[1]{Section~\ref{#1}}

\newcommand{\Figref}[1]{Fig.~\ref{#1}}

\graphicspath{{figures/}}
\numberwithin{equation}{section}

\begin{document}
\title{%
  Non-genericity of the Nariai solutions:\\ 
  I.\ Asymptotics and spatially homogeneous perturbations
}

\author{Florian Beyer\\\textit{\small beyer@ann.jussieu.fr}}
  
\date{{\small Laboratoire Jacques-Louis Lions\\
Universit\'e Pierre et Marie Curie (Paris 6)\\
4 Place Jussieu, 75252 Paris, France}}

\maketitle 

\begin{abstract}
This is the first of two papers where we study the asymptotics of the
generalized Nariai solutions and its relation to the cosmic no-hair
conjecture.  According to the cosmic no-hair conjecture, generic
expanding solutions of Einstein's field equations in vacuum with a
positive cosmological constant isotropize and approach the de-Sitter
solution asymptotically. The family of solutions which we introduce as
``generalized Nariai solutions'', however, shows quite unusual
asymptotics and hence should be non-generic in some sense. In this
paper, we list basic facts for the Nariai solutions and characterize
their asymptotic behavior geometrically. One particular result is a
rigorous proof of the fact that the Nariai solutions do not possess
smooth conformal boundaries.  We proceed by explaining the
non-genericity within the class of spatially homogeneous solutions. It
turns out that perturbations of the three isometry classes of
generalized Nariai solutions are related to different mass regimes of
Schwarzschild de-Sitter solutions. A motivation for our work here is
to prepare the second paper devoted to the study of the instability of
the Nariai solutions for Gowdy symmetry.  We will be particularly
interested in the construction of new and in principle arbitrarily
complicated cosmological black hole solutions.

%%% Local Variables: 
%%% mode: latex
%%% TeX-master: "paper"
%%% End: 

\end{abstract}

\renewcommand{\figurename}{Fig.}
\bibliographystyle{hplain}

\section{Introduction}
The general knowledge of qualitative properties and phenomena for
cosmological solutions of Einstein's field equations is still quite
limited. Of particular fundamental importance are the strong cosmic
censorship and the so-called BKL-conjectures.  The reader can find
relevant background material in \cite{andersson04a}. Recently, there
have been several break-throughs, but still many issues are open due
to the complexity of the interplay of geometry and highly complicated
partial differential equations in Einstein's theory.  In this paper,
another fundamental open problem will be the main concern. In a
restricted setting, we will study the asymptotic behavior of rapidly
expanding solutions of the field equations.  

A deeper understanding of such solutions is crucial for the correct
interpretation of cosmological observations.  The standard model of
cosmology, whose details are explained extensively e.g.\ in
\cite{Mukhanov06}, is surprisingly consistent with current
observations \cite{Sanchez06,Spergel06}.  A keystone of the standard
model is inflation in the very early universe, i.e.\ a period
characterized by rapid expansion. The theory of inflation has mostly
been formulated in the context of spatially homogeneous and isotropic
models. However, strong inhomogeneities can be expected in the very
early universe caused by quantum fluctuations and other physical
processes which are not yet completely understood.  One is interested
in the problem of whether inflation is able to homogenize and
isotropize the solutions despite possible strong inhomogeneities
before inflation. If so, this would yield a natural explanation for
the apparent homogeneity and isotropy of our universe without the
necessity of ad-hoc assumptions about the beginning of the universe.
Let us assume the simplest model for dark energy, namely a positive
cosmological constant $\Lambda>0$ in Einstein's field
equations. Recall that dark energy is the name for that hypothetical
matter component in our universe responsible for ``driving''
inflation. The conjecture that generic expanding solutions of
Einstein's field equations in vacuum with $\Lambda>0$,
\begin{equation}
  \label{eq:EFE}
  G_{\mu\nu}+\Lambda g_{\mu\nu}=0,
\end{equation}
approach the de-Sitter solution asymptotically is known as the cosmic
no-hair conjecture \cite{gibbons77,Hawking82}. Recall that the
de-Sitter solution is a homogeneous and isotropic inflationary
solution of \Eqref{eq:EFE}. If this conjecture were true, the
asymptotic state of generic solutions would be characterized by
homogenization, isotropization and exponential expansion.

Although there is some support for this conjecture in special
situations
\cite{Wald83,weber84,friedrich86,barrow88,Moniz,Kitada,Ringstrom06c},
the general case remains unclear due to the complexity of Einstein's
field equations. From this point of view, a particularly interesting
family of solutions of \Eqref{eq:EFE} is the class of Nariai solutions
which incorporates the (standard) Nariai solution
\cite{Nariai50,Nariai51}. Their relevance for our questions results
from the fact that they are \textit{simple} solutions which are
\textit{not} consistent with the cosmic no-hair picture. If the cosmic
no-hair conjecture is true, as is usually expected, this family of
solutions must be non-generic in a certain sense and in particular
unstable under arbitrary small perturbations. The perturbed solutions
should generically either collapse and form a singularity in a given
time direction, or if there is expansion, they should approach the
de-Sitter solution. The main question for this and the second paper
\cite{beyer09:Nariai2} is whether the expected instability of the
Nariai solutions under perturbations is true, and if so, how this is
realized dynamically.

Let us list some of the main assumptions. Throughout both papers, we
will restrict our attention to cosmological spacetimes by which we
mean $4$-dimensional globally hyperbolic Lorentzian manifolds with
compact spatial topology. It will turn out that spatial
\SoXSt-topology will be of particular interest for us. By cosmological
solutions we mean cosmological spacetimes which solve \Eqref{eq:EFE}
for $\Lambda>0$. This means that all our results are purely classical,
i.e.\ all quantum effects are ignored.  In general, when we speak of a
perturbation of a Nariai solution, we mean a cosmological solution of
the fully non-linear Einstein's field equations \Eqref{eq:EFE} whose
data, on some Cauchy surface, is close to the data on a Cauchy surface
of a given Nariai solution. By ``close'' we mean that two data sets
should deviate not too much with respect to some reasonable norm in
the initial data space. In principle, we are interested in generic
perturbations without symmetries. In practice, however, we must make
simplifying assumptions, and this paper focuses on the spatially
homogeneous case as a first step. In the second paper
\cite{beyer09:Nariai2}, we continue these investigations in a more
complicated, in particular inhomogeneous class of perturbations.

This paper is organized as follows. In \Sectionref{sec:conventions},
we collect our conventions and notation for both
papers. \Sectionref{sec:cosmicnohairnariai} contains a central part of
this paper. In \Sectionref{sec:cosmicnohair}, we give a short
discussion of the cosmic no-hair conjecture. We discuss conformal
boundaries and their relevance for our interests here.  Then,
in \Sectionref{sec:standardNariai}, we introduce the (standard) Nariai
solution and list its main properties. We analyze its asymptotics and
define ``Nariai asymptotics''. We discuss its relation to the cosmic
no-hair conjecture and prove that the Nariai solution has no smooth
conformal boundaries. Afterwards, we introduce the family of
generalized Nariai solutions
in \Sectionref{sec:generalizedNariai}. The standard Nariai solution
introduced earlier is a particular member of this family. We show that
all generalized Nariai solutions are locally isometric to the
(standard) Nariai solution, but in general have different global
properties. In any case, all of them have Nariai asymptotics at least
in one time direction. Although not all members in this family are
relevant for our interests here by themselves, there is, first, an
interesting property of their spatially homogeneous perturbations, as
we find in \Sectionref{sec:KS}, and, second, it will turn out that the
standard Nariai solution is not sufficient for the particular approach
taken in the second paper.  In the following, when we speak of a
``Nariai solution'', we mean a generalized Nariai solution.  The next
central part of this paper is \Sectionref{sec:KS}. We discuss
spatially homogeneous solutions of \Eqref{eq:EFE} for spatial
\SoXSt-topology, also known as the Kantowski-Sachs family, close to
(and including) the Nariai solutions. The cosmic no-hair conjecture
has been studied before in this family in
\cite{weber84,weber87,barrow88,Moniz}. We rederive those results which
are of relevance for the subsequent work and make some new
observations.  The precise description of the instability in the
spatially homogeneous case is of relevance for our second paper
\cite{beyer09:Nariai2}. The paper is concluded with a summary and a
short outlook to the second paper in \Sectionref{sec:summary}.

%%% Local Variables: 
%%% mode: latex
%%% TeX-master: "paper"
%%% End: 

\section{Notation and conventions}
\label{sec:conventions}
Throughout this and the second paper \cite{beyer09:Nariai2}, we assume
Einstein's summation convention when we write tensorial
expressions. Any tensor will either be represented by the abstract
symbol, say, $T$ or by abstract index notation, e.g.\
$T\indices{^\mu_\nu}$, depending on the context. Note, however, that
when we write such an indexed object, we can mean either the abstract
tensor or a special component with respect to some basis. Hopefully,
this will always be clear from the context. Our convention is that
Greek indices $\mu,\nu,\ldots=0,\ldots,3$ refer to some choice of
local spacetime coordinates, whereas Latin indices
$i,j,\ldots=0,\ldots 3$ represent indices with respect to some
orthonormal frame field. When we have chosen a time coordinate $t=x^0$,
then Greek indices $\alpha,\beta,\ldots=1,2,3$ represent spatial
coordinates, and for a choice of frame $\{e_i\}$ with timelike vector
field $e_0$, Latin indices $a,b,\ldots=1,2,3$ refer to spatial frame
indices. Writing $\{e_i\}$ just means the collection of tangent vector
fields $\{e_0,\ldots,e_3\}$. If a $2$-indexed object is written in
brackets, for instance $(T\indices{^\mu_\nu})$, we mean the matrix of
its components, where the first index labels the lines of the matrix
and the second one the columns.
%%% Local Variables: 
%%% mode: latex
%%% TeX-master: "paper"
%%% End: 

\section{Cosmic no-hair and the Nariai spacetimes}
\label{sec:cosmicnohairnariai}

\subsection{The cosmic no-hair conjecture}
\label{sec:cosmicnohair}
%\subsubsection{The conjecture and some problems}
\begin{Conj}[Cosmic no-hair conjecture]
  %\label{conj:Cosmicnohair}
  Any generic future causal geodesically complete asymptotically
  future expanding cosmological solution of the vacuum Einstein's
  field equations with $\Lambda>0$ is foliated by Cauchy surfaces
  which approach a homogeneous and isotropic foliation of the
  de-Sitter solution locally asymptotically to the future.
\end{Conj}
We say that a future causal geodesically complete cosmological
solution ``expands asymptotically to the future'' provided there is a
foliation of Cauchy surfaces labeled by time $t$ whose mean curvature
defined with respect to the future directed timelike unit normal is
strictly positive for large $t$. When the solution is not necessarily
future causal geodesically complete, but there is a foliation with
strictly negative mean curvature on a relevant region of the maximal
globally hyperbolic extension, we say that the solution ``collapses''.
Note that these definitions rule out cosmological black hole
spacetimes, for example the Schwarzschild-de-Sitter solution below,
which are characterized by the existence of both collapsing and
expanding regions. By a ``homogeneous and isotropic foliation of the
de-Sitter solution'', we mean a family of Cauchy surfaces forming a
foliation in the de-Sitter spacetime with a smooth transitive
isometric action of a $6$-dimensional Lie group on each leaf. The main
example is given by the standard $t=const$-surfaces in the de-Sitter
solution.  In the following, we will often speak of a solution
satisfying the ``cosmic no-hair picture'', when we mean that the
solution has a foliation with the asymptotic properties of the
conjecture.

It is difficult to find a precise formulation of this conjecture for
several reasons. Some problems are listed in the introduction of
\cite{Wald83}. We just note that the precise definitions of the
notions of ``genericity'' and ``approach'' are not fixed by the
conjecture. Apart from this, it is often difficult in practice to
identify a foliation with the properties above for a given solution.
In the spatially homogeneous case, there are geometrically preferred
foliations and the analysis can simplify. Indeed, there are several
results in the literature case which yield conditions in the spatially
homogeneous so that \textit{homogeneous} foliations approach a
homogeneous and isotropic foliation of the de-Sitter solution. The
first theorem in this direction was found in \cite{Wald83} and is
valid for all Bianchi solutions of Einstein's field equations with
$\Lambda>0$ and matter under certain conditions. Other results in the
spatially homogeneous case are
\cite{weber84,weber87,barrow88,Moniz,Kitada}. Further fundamentally
important results in this context are the non-linear stability of the
de-Sitter solution \cite{friedrich86,Lubbe09} and the stability theorem in
\cite{Ringstrom06c}, which both imply the cosmic no-hair conjecture
close to certain reference solutions without symmetry assumptions.

%\subsubsection{Relation between cosmic no-hair and smooth conformal
%  boundaries}
Since in practice it can be difficult to check whether a given
spacetime satisfies the cosmic no-hair picture, we introduce a
particularly important, sometimes simpler but less general, condition
now.  In this and the following paper \cite{beyer09:Nariai2}, this
condition shows up naturally and is used.  One way of discussing
asymptotics of spacetimes geometrically is to study the existence and
properties of conformal boundaries. Let us consider the class of
solutions of Einstein's field equations in vacuum with $\Lambda>0$,
which develop smooth future conformal boundaries \scrip
\cite{Friedrich2002, DeSitter, galloway2002}, for the following
paragraph. The analogous notion for the past time direction
exists. These solutions are also called future asymptotically
de-Sitter for a reason to become clear in a moment. Smooth conformal
boundaries correspond to the infinite timelike future or past in our
setting and hence their properties characterize the timelike
asymptotics of the spacetimes.  
%%%%%%%%%%%%%%%%%%%%%%%%%%%%%%%%%%%%%%%%%%%%%%%%%%%%%%%%%%%%%%%%%%%%% 
%%Short version of the article                                     %%
%%%%%%%%%%%%%%%%%%%%%%%%%%%%%%%%%%%%%%%%%%%%%%%%%%%%%%%%%%%%%%%%%%%%%
\opt{shortversion}{%
  The conformal formalism is particularly useful in the context of our
  two papers due to the following result which we state without proof.
}%
%%%%%%%%%%%%%%%%%%%%%%%%%%%%%%%%%%%%%%%%%%%%%%%%%%%%%%%%%%%%%%%%%%%%%
%%%%%%%%%%%%%%%%%%%%%%%%%%%%%%%%%%%%%%%%%%%%%%%%%%%%%%%%%%%%%%%%%%%%% 
%%Long version of the article                                      %%
%%%%%%%%%%%%%%%%%%%%%%%%%%%%%%%%%%%%%%%%%%%%%%%%%%%%%%%%%%%%%%%%%%%%%
\opt{longversion}{%
  The conformal formalism is particularly useful in the context of our
  two papers due to the following results.  
}%
%%%%%%%%%%%%%%%%%%%%%%%%%%%%%%%%%%%%%%%%%%%%%%%%%%%%%%%%%%%%%%%%%%%%%
\begin{Prop}
  \label{prop:confbound_CNH}
  Let a future causal geodesically complete cosmological solution for
  $\Lambda>0$ be given. Suppose it has a non-empty future conformal
  boundary $\scrip$.  Then, in a neighborhood of \scrip, there exists
  a local foliation of spacelike surfaces which approaches a
  homogeneous and isotropic foliation of the de-Sitter solution
  asymptotically.
\end{Prop}
A particular example is the de-Sitter solution itself. By ``approach''
we mean pointwise convergence along each timeline at $\scrip$.
%%%%%%%%%%%%%%%%%%%%%%%%%%%%%%%%%%%%%%%%%%%%%%%%%%%%%%%%%%%%%%%%%%%%% 
%%Long version of the article                                      %%
%%%%%%%%%%%%%%%%%%%%%%%%%%%%%%%%%%%%%%%%%%%%%%%%%%%%%%%%%%%%%%%%%%%%%
\opt{longversion}{%
  For the proof, let $(M,g)$ be spacetime under consideration and
  $(\tilde M,\tilde g,\Omega)$ its smooth conformal
  extension\footnote{Note that we use the opposite notation to
    \cite{beyer09:Nariai2}. In our work here, all quantities related to
    the conformal metric $\tilde g$ are marked with a tilde, while all
    quantities related to the physical metric $g$ have no mark.}. We
  identify $M$ with the interior of $\tilde M$.  The conformal factor
  $\Omega$ is a smooth function which is strictly positive on $M$, so
  that $\Omega=0$ and $d\Omega\not=0$ on $\scrip$. The identity
  $g=\Omega^{-2}\tilde g$ holds on $M$. Because $\scrip$ is spacelike
  for solutions of the vacuum field equations with $\Lambda>0$
  \cite{Friedrich2002}, we can foliate a neighborhood $U$ in $\tilde M$
  by spacelike level sets of $\Omega$. Let the future directed timelike
  unit normal tangent vector field of these surfaces with respect to
  $\tilde g$ be $\tilde N$, and with respect to $g$ be $N$. Note that
  $N=\Omega\tilde N$, and that both $N$ and $\tilde N$ are smooth on $U$
  with $\tilde N$ non-vanishing.  We choose a smooth non-vanishing (not
  necessarily orthonormal) frame $\{b_i\}$ with $b_0=\tilde N$ and $b_a$
  ($a=1,2,3$) orthogonal to $\tilde N$ on $U$. In particular, all $b_a$
  are hence tangent to the level sets and spacelike.  The components of
  the 2nd fundamental form of the level sets with respect to $g$ and
  $\tilde g$, respectively, for the frame $\{b_i\}$ are
  \begin{equation*}
    \chi_{ij}:=g(N, b_i)g(\nabla_{N}N, b_j)+g(\nabla_{b_i}N, b_j),\quad
    \tilde\chi_{ij}:=\tilde g(\tilde N, b_i)
    \tilde g(\tilde\nabla_{\tilde N}\tilde N, b_j)
    +\tilde g(\tilde\nabla_{b_i}\tilde N, b_j).
  \end{equation*}
  Here, $\nabla$ and $\tilde\nabla$ are the Levi-Civita covariant
  derivative operators of $g$ and $\tilde g$ respectively. It is a
  standard fact that $\chi_{0i}=0$, $\chi_{[ij]}=0$, and the same for
  $\tilde\chi_{ij}$. Moreover, one can check that
  \[\chi_{ij}=\Omega^{-1}\tilde\chi_{ij}
  -\Omega^{-2}\tilde N(\Omega) \tilde h_{ij},
  \] 
  where $\tilde h_{ij}$ are the components of the metric induced on the
  level sets by $\tilde g$. The following decomposition is standard
  \[\chi_{ij}=H h_{ij}+\sigma_{ij},\]
  where the expansion or mean curvature $H$ is three times the trace of
  $\chi_{ij}$ with respect to $g$ and the shear tensor $\sigma_{ij}$ is
  the traceless part.  The same is done for $\tilde\chi_{ij}$ with
  respect to $\tilde g$ yielding quantities $\tilde H$ and
  $\tilde\sigma_{ij}$.  It follows that
  \[H=\Omega\tilde H-\tilde N(\Omega),\quad
  \sigma\indices{^i_j}=\Omega\tilde\sigma\indices{^i_j}.
  \] 
  In the latter expression, the index has been raised with $g$ for the
  quantity $\sigma\indices{^i_j}$ and for $\tilde\sigma\indices{^i_j}$
  with $\tilde g$.  Now, since the conformal extension $(\tilde M,\tilde
  g,\Omega)$ and the frame $\{b_i\}$ is smooth on the conformal
  boundary, the quantities $\tilde H$, and $\tilde\sigma\indices{^i_j}$
  take finite values on $\scrip$. Moreover, Einstein's field equations
  in vacuum with $\Lambda>0$ imply \cite{DeSitter}
  \[\left.-\tilde N(\Omega)\right|_{\scrip}=\sqrt{\frac\Lambda 3}.\]
  Hence, we see that in a neighborhood of $\scrip$ (given by
  $\Omega=0$),
  \[H=\sqrt{\frac\Lambda 3}+O(\Omega),
  \quad
  \sigma\indices{^i_j}=O(\Omega).
  \]
  The value of $H$ is hence consistent with the expansion of the
  de-Sitter solution and the shear vanishes asymptotically at $\scrip$;
  note that this is the case in particular when $\{b_i\}$ is substituted
  by a frame which is either orthonormal with respect to $g$ or $\tilde
  g$. Now, concerning the curvature of the spatial slices, we find the
  following. Since in general the Ricci tensor is invariant under
  conformal rescalings of the metric with constant conformal factors,
  the Ricci tensor $r$ of the spatial metric $h$ equals the Ricci tensor
  $\tilde r$ of the spatial metric $\tilde h$. Furthermore, the frame
  components $\tilde r$ for an orthonormal frame of $\tilde h$ have
  finite values on $\scrip$ by virtue of the same argument as
  before. This shows that the components of $r$ with respect to an
  orthonormal frame of $h$ must be $O(\Omega^2)$ close to $\scrip$. Thus
  the physical curvature of the slices approaches zero. In the same way
  as \cite{Wald83}, we interpret this as the local approach to the
  de-Sitter solution which completes the proof.
  
  We remark that it is a priori not clear whether the surfaces in the
  proof of the previous proposition can be extended to surfaces which
  are both level sets of $\Omega$ and Cauchy surfaces globally. If
  $\scrip$ is compact, which is the most important case in the
  following, this extension, however, is possible.

  For later convenience, another notion \cite{Ringstrom06c} is useful in
  this context.  One says that future late time observers in a future
  causal geodesically complete cosmological spacetime are ``completely
  oblivious to topology'' if there is a Cauchy surface $\Sigma$ such
  that for all future inextendible causal curves $\gamma$, the
  intersection of $\Sigma$ with the causal past of $\gamma$ is contained
  in a subset of $\Sigma$ homeomorphic to a Euclidean $3$-ball. In this
  situation, observers have no information about the manifold properties
  of $\Sigma$ whatsoever because they can only see a ``topologically
  trivial'' $3$-ball at most. Note that the assumption of future causal
  geodesic completeness rules out cosmological black hole
  solutions. Clearly, there is also a past dual of this notion.
  \begin{Prop}
    \label{prop:compactscri}
    For any future causal geodesically complete cosmological spacetime
    $(M,g)$ with a spacelike smooth compact future conformal boundary,
    late time observers are completely oblivious to topology.
  \end{Prop}
  In order to prove this proposition, recall that due to the
  assumptions, the future conformal boundary \scrip is a Cauchy surface
  \cite{galloway2002} of the conformal extension $(\tilde M,\tilde
  g,\Omega)$ of $(M, g)$. In particular $\scrip$ is homeomorphic to all
  Cauchy surfaces of $(M,g)$. This implies that any given future
  directed inextendible causal curve of $(M,g)$ can be extended in
  $(\tilde M,\tilde g)$ to hit $\scrip$. Moreover, there is a
  neighborhood of \scrip in $(\tilde M,\tilde g)$ with a Gaussian time
  function $\tau$ with respect to \scrip such that the
  $\tau=0$-hypersurface equals $\scrip$ and all $\tau=const$-surfaces
  with $\tau<0$ are Cauchy surfaces of $(M, g)$. Now if we choose a
  $\tau_0<0$ with $|\tau_0|$ small enough, then the
  $\tau=\tau_0$-surface $\Sigma_0$ has the required properties. That is,
  choose any future directed inextendible causal curve $\gamma$ on
  $(M,g)$ with future endpoint $p\in\scrip$. Choose a neighborhood $V$
  of $p$ in $\scrip$ homeomorphic to a $3$-ball. Let $V_0$ be the set in
  $\Sigma_0$ obtained from $U$ via the past flow of the Gaussian time
  vector field. Then the causal past of $p$ intersected with $\Sigma_0$
  is contained in $V_0$ if $|\tau_0|$ is sufficiently small. Now the
  proposition follows because the causal past of $p$ comprises the
  causal past of $\gamma$. %We remark that some of the strong assumptions
  % made here can possibly be relaxed, but we will not elaborate on this.
  
  Since the de-Sitter solution possesses a smooth compact \scrip,
  \Propref{prop:compactscri} implies that late time observers are
  completely oblivious to topology. If a solution obeys the cosmic
  no-hair picture with respect to some foliation and hence approaches
  the de-Sitter solution in a ``sufficiently well behaved'' manner to
  the future, one can expect that it has the same property. Based on
  this argument, we conjecture the following alternative to the cosmic
  no-hair conjecture: For generic future causal geodesically complete
  future asymptotically expanding cosmological solutions of the vacuum
  field equations with $\Lambda>0$, future late time observers are
  completely oblivious to topology.
}%
%%%%%%%%%%%%%%%%%%%%%%%%%%%%%%%%%%%%%%%%%%%%%%%%%%%%%%%%%%%%%%%%%%%%%

%%% Local Variables: 
%%% mode: latex
%%% TeX-master: "paper"
%%% End: 

\subsection{Standard Nariai spacetime}
\label{sec:standardNariai}
\subsubsection{Basic properties}
Let $\Lambda$ be an arbitrary number. Consider the Lorentzian manifold
$(M,g)$ given by
\begin{equation}
  \label{eq:StandardNariai}
  M=\R\times(\SoXSt),\quad g=\frac 1\Lambda\left(
    -dt^2+\cosh^2t\,d\rho^2+g_{\St}\right),
\end{equation}
with the standard coordinate $\rho$ on the manifold $\So$, the
standard round unit metric $g_{\St}$ on $\St$ and the time coordinate
$t\in\R$. This spacetime is called the Nariai spacetime. It was first
discussed by Nariai \cite{Nariai50,Nariai51} and later reconsidered in
various works; an overview of references is given by
\cite{Bousso03}. It is also sometimes called Weber solution due to the
work in \cite{weber84,weber87}.  This solution has turned out to be
very interesting from the semi-classical point of view because it has
certain extremal horizon properties \cite{Bousso03}. We will not
discuss such issues here.  Since we also want to consider
``generalized'' Nariai spacetimes
in \Sectionref{sec:generalizedNariai}, we often speak of the
``standard'' Nariai spacetime where necessary. We will also often
consider the universal cover of the standard Nariai spacetime. This is
the simply connected Lorentzian manifold $(\tilde M,g)$ with $\tilde
M=\R\times(\R\times\St)$ and the metric $g$ above.

The standard Nariai spacetime has the following properties.
\renewcommand{\labelenumi}{(\roman{enumi})}
\begin{enumerate}
\item It is an analytic cosmological solution of Einstein's field
  equations in vacuum with a cosmological constant $\Lambda>0$. Hence
  we will often also speak of the (standard) Nariai
  \textit{solution}. The $t=const$-hypersurfaces are spacelike Cauchy
  surfaces of topology \SoXSt. 
\item It is geodesically complete because it is the direct product of
  the $2$-dimensional de-Sitter spacetime and a rescaled round
  $2$-sphere.
\item The representation \Eqref{eq:StandardNariai} of the universal
  cover of the standard Nariai spacetime is the maximal analytic
  extension. Maximality follows from geodesic completeness. Uniqueness
  is implied by the version of Theorem~6.3 in \cite{kobayashi} for
  Lorentzian metrics, or by Theorem~4.6 from \cite{Chrusciel08} for
  the special case of geodesic completeness. In the literature one
  finds other representations of the Nariai metric, see the references
  above, which are not always geodesically complete and hence not
  necessarily maximal.
\item It is spatially homogeneous with Kantowski-Sachs symmetry group
  $G=\R\times\SO$. 
\item The Nariai solution $(M,g)$ has ``Nariai asymptotics'';
  a term which we explain now.
\end{enumerate}

\subsubsection{Nariai asymptotics and cosmic no-hair}
A particularly peculiar property of the Nariai solution is its time
dependence. While the \So-factor of the spatial slices expands
exponentially for increasing positive $t$, the volume of the
\St-factor stays constant. Thus, the expansion of this solution is
anisotropic in the sense that the shear tensor of the $t=const$-slices
never approaches zero%
%%%%%%%%%%%%%%%%%%%%%%%%%%%%%%%%%%%%%%%%%%%%%%%%%%%%%%%%%%%%%%%%%%%%% 
%%Long version of the article                                      %%
%%%%%%%%%%%%%%%%%%%%%%%%%%%%%%%%%%%%%%%%%%%%%%%%%%%%%%%%%%%%%%%%%%%%%
\opt{longversion}{%
  ; cf.\ the proof of \Propref{prop:confbound_CNH}%
}. %
%%%%%%%%%%%%%%%%%%%%%%%%%%%%%%%%%%%%%%%%%%%%%%%%%%%%%%%%%%%%%%%%%%%%%
This implies that the cosmic no-hair picture does not hold with
respect to the foliation of $t=const$-surfaces. So we must ask whether
there exist other foliations of the Nariai solution, presumably based
on non-symmetric surfaces, for which the cosmic no-hair picture is
attained. In this paper, we are not able to answer this question
completely, and we stress that to our knowledge there is no result in
the literature which implies that the Nariai solution contradicts the
cosmic no-hair picture for \textit{all} foliations.  
%%%%%%%%%%%%%%%%%%%%%%%%%%%%%%%%%%%%%%%%%%%%%%%%%%%%%%%%%%%%%%%%%%%%% 
%%Long version of the article                                      %%
%%%%%%%%%%%%%%%%%%%%%%%%%%%%%%%%%%%%%%%%%%%%%%%%%%%%%%%%%%%%%%%%%%%%%
\opt{longversion}{%
  Nevertheless, we discuss a property of this solution now which
  suggests that the cosmic no-hair picture does not hold for the
  Nariai solution.  Then we prove a weaker statement below, to the
  first time rigorously in the literature, that the Nariai solution
  does not possess smooth conformal boundaries.  
}%
%%%%%%%%%%%%%%%%%%%%%%%%%%%%%%%%%%%%%%%%%%%%%%%%%%%%%%%%%%%%%%%%%%%%%
%%%%%%%%%%%%%%%%%%%%%%%%%%%%%%%%%%%%%%%%%%%%%%%%%%%%%%%%%%%%%%%%%%%%% 
%%Short version of the article                                     %%
%%%%%%%%%%%%%%%%%%%%%%%%%%%%%%%%%%%%%%%%%%%%%%%%%%%%%%%%%%%%%%%%%%%%%
\opt{shortversion}{%
  In any case, we show now that the Nariai solution does not possess
  smooth conformal boundaries, and according to our discussions before,
  this gives reasons to believe that the cosmic no-hair picture does
  not hold.  
}%
%%%%%%%%%%%%%%%%%%%%%%%%%%%%%%%%%%%%%%%%%%%%%%%%%%%%%%%%%%%%%%%%%%%%%

%%%%%%%%%%%%%%%%%%%%%%%%%%%%%%%%%%%%%%%%%%%%%%%%%%%%%%%%%%%%%%%%%%%%% 
%%Long version of the article                                      %%
%%%%%%%%%%%%%%%%%%%%%%%%%%%%%%%%%%%%%%%%%%%%%%%%%%%%%%%%%%%%%%%%%%%%%
\opt{longversion}{%
  Ringström \cite{Ringstrom06c} introduces the terminology that in a
  future causal geodesically complete spacetime, future late time
  observers are ``oblivious to topology'' if there is a Cauchy surface
  $\Sigma$ such that there are no future directed inextendible causal
  curves whose causal pasts contain $\Sigma$. The reader should
  distinguish this from the notion ``completely oblivious to topology''
  introduced in \Sectionref{sec:cosmicnohair}.  Again, there clearly
  exists a corresponding past dual of this notion, but we restrict to
  the future. In physical terms this means that all light rays are
  confined to a subregion of the spatial slices due to the rapid
  expansion.

  It is evident that for spacetimes whose observers are completely
  oblivious to topology, observers are also oblivious to topology. In
  particular, for the de-Sitter solution it thus follows that observers
  are oblivious to topology, and indeed also for the Nariai
  solution. However, the Nariai solution is an example with the
  following property.
  \begin{Prop}
    For the Nariai solution, and also for its universal cover, future
    late time observers are oblivious to topology, but not completely
    oblivious to topology.
  \end{Prop}
  In the following, we say that all solutions with this property have
  ``Nariai asymptotics''.  This proposition is a consequence of an even
  stronger result by Ringström \cite{Ringstrom06c}.
}%
%%%%%%%%%%%%%%%%%%%%%%%%%%%%%%%%%%%%%%%%%%%%%%%%%%%%%%%%%%%%%%%%%%%%%
%%%%%%%%%%%%%%%%%%%%%%%%%%%%%%%%%%%%%%%%%%%%%%%%%%%%%%%%%%%%%%%%%%%%% 
%%Short version of the article                                     %%
%%%%%%%%%%%%%%%%%%%%%%%%%%%%%%%%%%%%%%%%%%%%%%%%%%%%%%%%%%%%%%%%%%%%%
\opt{shortversion}{%
  Ringström \cite{Ringstrom06c} shows the following property
  which we refer to as ``Nariai asymptotics'' in the following.  
}%
%%%%%%%%%%%%%%%%%%%%%%%%%%%%%%%%%%%%%%%%%%%%%%%%%%%%%%%%%%%%%%%%%%%%%
That is, regardless of the choice of Cauchy surface $\Sigma$ and
future directed inextendible causal curve $\gamma$ in the Nariai
solution, the intersection of $\Sigma$ and the causal past of $\gamma$
is not contained in a set homeomorphic to a $3$-ball. In simple words,
although observers in the Nariai solutions do not have full
information about the spatial topology due to the rapid expansion,
they will always have \textit{some} information because the anisotropy
of the expansion allows to see the full \St-factor
asymptotically. Ringström's original proof of this statement in
Lemma~21 in \cite{Ringstrom06c} uses the compactness of the Cauchy
surfaces of the standard Nariai solution and hence does not apply to
the universal cover. However, his argument can be rescued for the
universal cover as follows. Let $\Sigma$ be a Cauchy surface and
$\gamma$ a future directed inextendible causal curve in the universal
cover spacetime. Now the main new observation is that the closure of
$J^-(\gamma)\cap\Sigma$ is compact for the Nariai metric. Hence, when
we assume that this set is contained in $B\subset\Sigma$ homeomorphic
to a $3$-ball, we can assume in particular that $\bar B$ is
compact. Ringström's contradiction argument at the end of his original
proof can now be applied with respect to a value of $\tau$ chosen
large enough so that the $t=\tau$-surface $S_\tau$ is strictly to the
future of $\bar B$.

A particular consequence of Ringström's result above is that the
Nariai solution does not possess a smooth conformal boundary. This has
been conjectured in \cite{gibbons77} and in many references
afterwards. The conjecture was based on the simplest conformal
rescaling of the metric \Eqref{eq:StandardNariai} by means of a
conformal factor only depending on $t$ because this does not lead to a
smooth conformal compactification. However, in order to turn this
statement into a theorem, one needs to show that there exists no
conformal rescaling at all which would yield a smooth conformal
compactification. 
\begin{Prop}
  \label{prop:nariaiconfboundary}
  The standard Nariai solution does not have even a patch of a smooth
  conformal boundary. The same is true for its universal cover.
\end{Prop}
Suppose the contrary is the case, and an open patch of a smooth future
conformal boundary $\scrip$, not necessarily compact, exists for the
Nariai solution $(M,g)$. Since $(M,g)$ is a cosmological solution of
\Eqref{eq:EFE}, $\scrip$ is spacelike \cite{Friedrich2002}.  Then
there exists a future directed inextendible causal curve $\gamma$ of
$(M,g)$ which can be extended in the conformally extended spacetime
$(\tilde M,\tilde g)$ so that it hits \scrip in the point $p$. Let $V$
be a neighborhood of $p$ in $\scrip$ homeomorphic to a $3$-ball; this
neighborhood exists thanks to our assumption that \scrip consists at
least of an open patch. There is a neighborhood $U$ of $V$ in $(\tilde
M,\tilde g)$ with a Gaussian time function $\tau$ such that the
$\tau=0$-hypersurface equals $V$ and all $\tau=const$-surfaces with
$\tau<0$ are spacelike surfaces of $(M, g)$. Now, if we choose
$\tau_0<0$ with $|\tau_0|$ small enough, then any future directed
inextendible causal curve with $p$ as its future endpoint must
intersect the $\tau=\tau_0$-hypersurface. Since this surface is
homeomorphic to a $3$-ball, because $V$ has this property, and since
it can be extended to a Cauchy surface, because it is spacelike, we
have found a contradiction to Ringström's statement above, that no
Cauchy surface and causal curve with this property exist in the Nariai
solution. Hence, our claim is proven.

%%%%%%%%%%%%%%%%%%%%%%%%%%%%%%%%%%%%%%%%%%%%%%%%%%%%%%%%%%%%%%%%%%%%% 
%%Long version of the article                                      %%
%%%%%%%%%%%%%%%%%%%%%%%%%%%%%%%%%%%%%%%%%%%%%%%%%%%%%%%%%%%%%%%%%%%%%
\opt{longversion}{%
  According to our speculation about cosmic no-hair at the end
  of \Sectionref{sec:cosmicnohair}, all this suggests that the Nariai
  solution does not satisfy the cosmic no-hair picture and hence
  motivates our claim that it should be non-generic among solutions of
  the field equations.  
}%
%%%%%%%%%%%%%%%%%%%%%%%%%%%%%%%%%%%%%%%%%%%%%%%%%%%%%%%%%%%%%%%%%%%%%
%%%%%%%%%%%%%%%%%%%%%%%%%%%%%%%%%%%%%%%%%%%%%%%%%%%%%%%%%%%%%%%%%%%%% 
%%Short version of the article                                     %%
%%%%%%%%%%%%%%%%%%%%%%%%%%%%%%%%%%%%%%%%%%%%%%%%%%%%%%%%%%%%%%%%%%%%%
\opt{shortversion}{%
  By virtue of \Propref{prop:confbound_CNH}, this result suggests that
  the Nariai solution does not satisfy the cosmic no-hair picture and
  hence motivates our claim that it should be non-generic among
  solutions of the field equations.  
}%
%%%%%%%%%%%%%%%%%%%%%%%%%%%%%%%%%%%%%%%%%%%%%%%%%%%%%%%%%%%%%%%%%%%%%

%%% Local Variables: 
%%% mode: latex
%%% TeX-master: "paper"
%%% End: 

\subsection{Generalized Nariai spacetimes}
\label{sec:generalizedNariai}
\subsubsection{Definition and general properties}
Consider the Lorentzian manifold $(M,g)$ with
\begin{subequations}
  \label{eq:generalizedNariai}
  \begin{equation}
    g=\frac 1\Lambda\Bigl( -dt^2 +\Phi(t)^2\,d\rho^2+g_{\St}\Bigr)
  \end{equation}
  with
  \begin{equation}
    \Phi(t)=\Phi_0 \cosh t+\Phi_0^\prime\sinh t
  \end{equation}
\end{subequations}
for arbitrary $\Lambda>0$, $\Phi_0>0$ and $\Phi_0^\prime\in\R$ and with
\[M=I\times(\SoXSt).\] Here, $I$ is the unique maximal connected open
interval in \R with $0\in I$ and $\left.\Phi\right|_{I}>0$. Note that
$\Phi$ can change its sign at most once. We will henceforth refer to
this family as the ``generalized Nariai spacetimes''. Clearly, for
$\Phi_0=1$ and $\Phi_0^\prime=0$, we recover the standard Nariai
solution. All generalized Nariai spacetimes are analytic solutions of
the vacuum Einstein's field equations with a cosmological constant
$\Lambda$. They are globally hyperbolic, but not necessarily causal
geodesically complete; as we discuss further now. They are spatially
homogeneous of Kantowski-Sachs type. We will also often consider the
universal cover. This is the simply connected Lorentzian manifold
$(\tilde M,g)$ with $\tilde M=\R\times(\R\times\St)$ and the metric
$g$ above. Note that our family of generalized Nariai solution is
related to, but distinct from the family with the same name defined in
\cite{chruciel01} in Eq.~(1.2). In our language, the latter family
assumes $\Phi_0^\prime=0$ but arbitrary signs of $\Lambda$, while our
family restricts to $\Lambda>0$ with arbitrary $\Phi_0^\prime$.

\subsubsection{Global behavior}
Let us define
\begin{equation}
  \label{eq:defsigma0}
  \sigma_0:=\Phi_0^2-(\Phi_0^{\prime})^2.%=\Phi^2(t)-(\partial_t \Phi)^2(t).
\end{equation}
One finds easily that $\Phi$ defined in \Eqref{eq:generalizedNariai}
has a real zero if and only if $\sigma_0<0$. There cannot be more than
one real zero. In particular, $I=\R$ if and only if $\sigma_0\ge
0$. Similarly, one shows that $\Phi$ has one and only one local minimum
if and only if $\sigma_0>0$. 

Let us consider the universal cover of a generalized Nariai solution
in order to avoid problems with identifications on the \So-factor in
the following arguments. Furthermore, let us fix once and for all a
value of $\Lambda$. The discussion above implies that if $\sigma_0>0$,
we can shift the time coordinate $t$ so that $t=0$ is the time of the
minimum of $\Phi$ and hence $\Phi_0^\prime=0$ and $\Phi_0>0$. Note
that $\sigma_0$ is invariant under time translation. Then, by
rescaling the $\rho$-coordinate we can choose $\Phi_0=1$. Hence the
universal cover of any generalized Nariai solutions with $\sigma_0>0$
is isometric to the universal cover of the standard Nariai solution
with $\sigma_0=1$. The isometry involved here is analytic.  For
$\sigma_0=0$, the function $\Phi$ has neither a zero nor a local
extremum, and $\Phi_0^\prime=\pm \Phi_0$. By a change of time
direction and rescaling of the $\rho$-coordinate, we can always
arrange that $\Phi_0^\prime=\Phi_0=1$. Hence, the universal cover of
any generalized Nariai solution with $\sigma_0=0$ is isometric to the
universal cover of the generalized Nariai solution with
$\Phi_0^\prime=\Phi_0=1$. The isometry is again analytic.  Finally,
consider the universal cover of a generalized Nariai solution with
$\sigma_0<0$. By a rescaling of the $\rho$-coordinate, we can choose
$\sigma_0=-1$. Now, since $\Phi:I\rightarrow\R^{>0}$ is surjective,
there is a time translation which gives for instance
$\Phi_0=(e^2-1)/2e$. Since $\sigma_0=-1$, we can transform $t\mapsto
-t$ if necessary in order to get $\Phi_0^\prime=(e^2+1)/2e$. Thus the
universal cover of any generalized Nariai solutions with $\sigma_0<0$
is isometric to the universal cover of the generalized Nariai solution
given by $\Phi_0=(e^2-1)/2e$, $\Phi_0^\prime=(e^2+1)/2e$, and the
isometry is analytic.

Hence, we have identified three analytic isometry classes of
generalized Nariai solutions given by the sign of the constant
$\sigma_0$. Are they distinct, or related for instance to the standard
Nariai solution in some way? For the case $\sigma_0>0$ we have already
shown that the universal cover of the corresponding generalized Nariai
solution is globally isometric to the universal cover of the standard
Nariai solution. Let us consider the case $\sigma_0<0$. Without loss
of generality we can assume that $\Phi_0=(e^2-1)/2e$,
$\Phi_0^\prime=(e^2+1)/2e$ and hence that $I=(-1,\infty)$. Let us
denote the universal cover of this generalized Nariai solution by
$(\hat M,\hat g)$ and its coordinates by $(\hat
t,\hat\rho,\hat\theta,\hat\phi)$. The universal cover of the standard
Nariai solution is referred to as $(M,g)$ with coordinates
$(t,\rho,\theta,\phi)$. Consider the map
\[\Psi_-:\hat M\rightarrow M, \quad
  (\hat t, \hat\rho, p)\mapsto 
  (t(\hat t, \hat\rho),\rho(\hat t, \hat\rho),p)\]
where $p$ denotes an arbitrary point on \St and where
\begin{align*}
  t(\hat t, \hat\rho)
  &=\text{arcsinh}\left(\sinh(\hat t+1)\sqrt{\sinh^2\hat\rho+1}\right),\\
  \rho(\hat t, \hat\rho)
  &=\text{arcsin}\left(
      \frac{\sinh (\hat t+1) \sinh \hat\rho}
      {\sqrt{\left(\sinh ^2\hat\rho+1\right)
      \sinh ^2(\hat t+1)+1}}\right)+\pi.
\end{align*}
One can check that the map $\Psi_-$ is an analytic embedding and,
considered as a map onto its image, an isometry. It is obvious that
the same statement would be true if we added other constants to
$\rho(\hat t, \hat\rho)$. Before we discuss this map further, let us
continue with the case $\sigma_0=0$. Let us use the same notation as
before in this case for the map
\[\Psi_0:\hat M\rightarrow M.\]
When we set
\begin{align*}
  t(\hat t, \hat\rho)
  &=\text{arcsinh}\left(\frac{e^{\hat t}
      \left(\hat\rho^2+1\right)-e^{-\hat t}}{2}\right)\\
  \rho(\hat t, \hat\rho)
  &=\begin{cases}
    -\text{arccos}\left(
      \frac{e^{-\hat t}-e^{\hat t} \left(\hat\rho^2+1\right)+2e^{\hat t}}
      {\sqrt{
          \left(e^{\hat t} \left(\hat\rho^2+1\right)-e^{-\hat t}\right)^2+4}
      }
    \right) &
    \hat\rho<0, \\
    0 & \hat\rho=0 \\
    \text{arccos}\left(
      \frac{e^{-\hat t}-e^{\hat t} \left(\hat\rho^2+1\right)+2e^{\hat t}}
      {\sqrt{
          \left(e^{\hat t} \left(\hat\rho^2+1\right)-e^{-\hat t}\right)^2+4}
      }
    \right) &
    \hat\rho>0,
  \end{cases},
\end{align*}
we find that $\Psi_0$ is also an analytic isometric embedding.

\begin{figure}[t]
  \centering
  \psfrag{I}[][][0.9]{III}
  \psfrag{II}[][][0.9]{II}
  \psfrag{tau/pi}[tr][br][0.9]{$\tau$}
  \psfrag{rho/pi}[][][0.9]{$\rho$}
  \psfrag{-1}[][][0.7]{$-\pi$}
  \psfrag{-0.5}[][][0.7]{$-\pi/2$}
  \psfrag{ 0}[][][0.7]{$0$}
  \psfrag{ 0.5}[][][0.7]{$\pi/2$}
  \psfrag{ 1}[][][0.7]{$\pi$}
  \psfrag{ 1.5}[][][0.7]{$3\pi/2$}
  \includegraphics[width=0.6\textwidth]{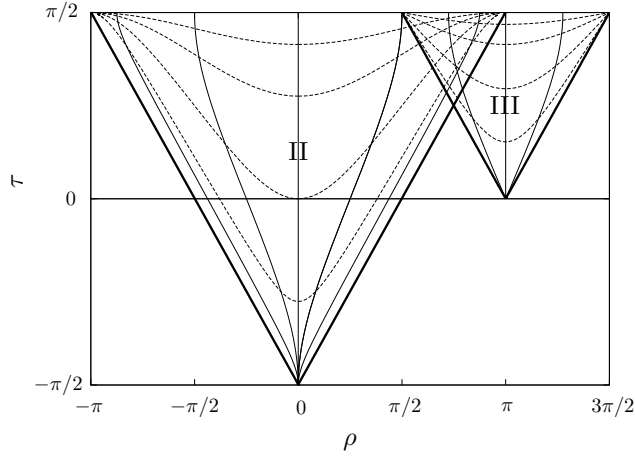}
  \caption{Embeddings of generalized Nariai solutions into the standard one}
  \label{fig:generalizedNariai}
\end{figure}
Hence we have shown for the universal covers, that all generalized
Nariai solutions can be embedded isometrically into the standard
Nariai solution. We visualize this in
\Figref{fig:generalizedNariai}. In this figure, the horizontal axis
represents the coordinate $\rho$ and the vertical axis shows
\[\tau:=2\arctan(\tanh(t/2)).\]
We consider again the universal cover of the standard Nariai solution.
This compactified time coordinate has the property that
$t\rightarrow\pm\infty$ corresponds to $\tau\rightarrow\pm\pi/2$. Each
point in this diagram determines a $2$-sphere in the standard Nariai
solution. In these coordinates, null curves which are constant on the
\St-factor are straight lines and their angles have the Minkowski
value.  Hence the picture appears like a Penrose diagram. However,
note that this would not be the case if we considered general null
curves.  For the case $\sigma_0>0$, the corresponding generalized
Nariai solution is isometric globally to the standard Nariai
solution. For $\sigma_0=0$, the region covered by the embedding
$\Psi_0$ is given by the region II in \Figref{fig:generalizedNariai}
for $\Phi_0^\prime>0$ which we can always assume. The thin dashed
curves represent $\hat t=const$-surfaces and the non-dashed thin
curves $\hat\rho=const$-surfaces. The black bold lines correspond to
the limit $\hat t\rightarrow-\infty$. For the case $\sigma_0<0$ and
$\Phi_0^\prime>0$, the region III is covered by the map $\Psi_-$. Here
the black bold lines correspond to the curves $\hat t\rightarrow -1$
for our standard choice of parameters.

The above implies that the maximal analytic extensions of the
universal covers of all generalized Nariai solutions are isometric to
the universal cover of the standard Nariai solution. Our discussion
furthermore proves that for $\sigma_0\le 0$, all generalized Nariai
solutions are causal geodesically incomplete. Finally, after a change
of time direction if necessary, all generalized Nariai solutions have
Nariai asymptotics to the future.

All these statements hold for the universal covers and yield globally
defined isometric embeddings. Returning to the original manifolds, we
see that the maps corresponding to $\Psi_0$ and $\Psi_-$ are not
smooth globally for $\sigma_0\le0$, however are still local
isometries.

%%% Local Variables: 
%%% mode: latex
%%% TeX-master: "paper"
%%% End: 

\section{Instability of Nariai asymptotics within the Kantowski-Sachs
  family}
\label{sec:KS}
\subsection{Ansatz and general solution}
\label{sec:ansatzKS}
In the following, we explain the non-genericity of the Nariai
solutions within the spatially homogeneous class of solutions.  We
start by first writing the general solution of the field equations
which comprises the generalized Nariai solutions. Then we define
spatially homogeneous perturbations and study their properties.

More precisely, we consider the class of spacetimes with a manifold
given by $M=\R\times(\SoXSt)$ and a metric $g$ which is invariant
under a global smooth effective action of the Kantowski-Sachs group,
whose orbits are spacelike Cauchy surfaces. This allows us to write
the metric in the form \cite{Wainwright}
\begin{equation}
  \label{eq:KSGauss}
  g=\frac 1\Lambda\left(-dt^2+\bar F(t) d\rho^2+\bar G(t) g_{\St}\right)
\end{equation}
with similar conventions as before. The functions $\bar F$ and $\bar
G$ play a similar role as the single scale factor $a$ in
Friedmann-Robertson-Walker solutions \cite{Wainwright}. We factor out
$1/\Lambda$ so that all quantities in the bracket are
dimensionless. Recall that $\Lambda$ has the dimension
$\text{Length}^{-2}$ in geometric units and that the coordinate
components of $g$ have dimension $\text{Length}^{2}$ for dimensionless
coordinates. In order to simplify the problem of dimensions, we will
always assume $\Lambda=1$ in the following. A metric $g$, which is a
solution of \Eqref{eq:EFE} with $\Lambda=1$, yields a metric $\hat
g=g/\hat\Lambda$ which solves \Eqref{eq:EFE} with
$\Lambda=\hat\Lambda>0$. This means that the choice $\Lambda=1$ means
no loss of generality, and hence allows us to restrict to
dimensionless quantities in the following.

The metric given by \Eqref{eq:KSGauss} is well-defined as long as
$\bar F$ and $\bar G$ are positive. However, in this representation,
it is difficult to discuss the limit when $\bar F$ or $\bar G$ become
zero. This is the reason why we prefer to work with
``Eddington-Finkelstein''-like coordinates $(s,\mu,\theta,\phi)$ on
$\R\times(\SoXSt)$, related to the above Gaussian coordinates by
\begin{equation*}
  %\label{eq:coordtrafo}
  \sqrt{\bar F(t)}\, dt=ds, \quad d\rho=1/\bar F(t)\, ds+d\mu,
\end{equation*}
as long as $\bar F$ is positive.
In these coordinate the metric takes the form
\[g=2dsd\mu+F(s)d\mu^2+G(s) g_{\St}\]
where 
\[F(s)=\bar F(t(s)), \quad G(s)=\bar G(t(s)).\] Note that $s$ is a
null coordinate. The symmetry hypersurfaces are the $t=const$-, or
equivalently the $s=const$-surfaces.  As before for the coordinate
$\rho$, the function $\mu$ is the coordinate on the $\So$-factor,
generated by the translation subgroup of the Kantowski-Sachs symmetry
group. The motivation for this choice of coordinates is that the
metric is well-behaved when $F$ changes sign, the curvature is bounded
as long as $F$ is smooth and $G>0$, and hence the metric can sometimes
be extended even where the symmetry hypersurfaces change their causal
character.

Let us assume in the following that the initial hypersurface for the
initial value problem of \Eqref{eq:EFE} under these assumptions is
given by $s=0$ and is spacelike, i.e.\ $F(0),G(0)>0$. It is clear that
an $s$-neighborhood of the initial hypersurface is globally hyperbolic
and that the $s=const$-hypersurfaces are spacelike compact Cauchy
surfaces. One easily computes the mean curvature of the
$s=const$-hypersurfaces
\begin{equation}
  \label{eq:meancurvatureKS}
  H(s)=\frac 16\left(\frac 1{\sqrt{F(s)}}\frac{dF}{ds}
    +2\frac{\sqrt{F(s)}}{G(s)}\frac{dG}{ds}\right),
\end{equation}
which holds as long as the surface is spacelike.

\begin{subequations}
  \label{eq:KS-EddingtonFinkelstein}
  In the coordinates $(s,\mu,\theta,\phi)$, Einstein's field
  equations in vacuum with $\Lambda=1$ reduce to
  \begin{align}
    0&=\ddot G(s)-\frac{\dot G^2(s)}{2G(s)}\\
    0&=\ddot F(s)+\frac{\dot F(s)\dot G(s)}{G(s)}-2\\
    \label{eq:KS-EddingtonFinkelstein-constraint}
    0&=1
    -\frac 1{G(s)}-\frac{\dot F(s)\dot G(s)}{2G(s)}
    -\frac{F(s)\dot G^2(s)}{4G^2(s)},
  \end{align}
\end{subequations}
where a dot represents derivatives with respect to $s$.  The initial
value problem for these equations is well-posed under our assumptions,
and for given data on the $s=0$-hypersurface satisfying the constraint
\Eqref{eq:KS-EddingtonFinkelstein-constraint}, it has a unique
solution maximally extended in the coordinates
$(s,\mu,\theta,\phi)$. 
%\begin{subequations}
  %\label{eq:solKSFG}
  Indeed, all solutions can be written explicitly as
  \begin{align*}
    F(s)&=F_*+\dot F_* s\frac{2}{{H^{(0)}_*}s+2}
    +s^2\frac{{H^{(0)}_*} s+6}{3{H^{(0)}_*} s+6},\\
    G(s)&=\frac{G_*}4 ({H^{(0)}_*}s+2)^2,
  \end{align*}
%\end{subequations}
with constants $G_*>0$, $F_*>0$, $\dot F_*$ and ${H^{(0)}_*}$ corresponding
to the data of $F$ and $G$ on the initial hypersurface by
\[F_*=F(0),\quad G_*=G(0), \quad \dot F_*=(\partial_s F)(0),\quad
{H^{(0)}_*}=(\partial_s G)(0)/G(0).\]

\subsection{Characterization of the solutions}
Consider arbitrary data consistent with the constraint for
${H^{(0)}_*}=0$. The corresponding solution -- denoted $(M,g)$ in the
following -- is
\[F(s)=F_*+\dot F_* s+s^2,\quad G(s)=1,\] with $F_*>0$ and $\dot F_*$
arbitrary.  We find that $(M,g)$ is an analytic extension of the
generalized Nariai solution -- denoted by $(\tilde M, \tilde g)$ for
now -- given by
\begin{equation}
  \label{eq:NariaiIdent}
  \Phi_0=\sqrt{F_*},\quad \Phi_0^\prime=\dot F_*/2.
\end{equation}
The corresponding analytic isometric embedding $\tilde M\rightarrow M$
is given by
\begin{align*}
  s(t)&=
  (\partial_t\Phi)(t)-\Phi_0^\prime,\\
  \mu(\rho,t)&=\rho-
  \begin{cases}
    \frac 1{\sqrt{\sigma_0}}\left(
      \arctan\frac{(\partial_t\Phi)(t)}{\sqrt{\sigma_0}}
      -\arctan\frac{\Phi_0^\prime}{\sqrt{\sigma_0}}
    \right) & \sigma_0>0,\\
    -\frac 1{(\partial_t\Phi)(t)}+\frac 1{\Phi_0^\prime} & \sigma_0=0,\\
    -\frac 1{\sqrt{|\sigma_0|}}\left(
      \arctanh\frac{(\partial_t\Phi)(t)}{\sqrt{\sigma_0}}
      -\arctanh\frac{\Phi_0^\prime}{\sqrt{\sigma_0}}
    \right) & \sigma_0<0.
  \end{cases}
\end{align*}
This is why we call data for the Kantowski-Sachs initial value problem
with ${H^{(0)}_*}=0$ Nariai data.  With this we find the following
further aspects for generalized Nariai solutions. In these discussions
we identify $(\tilde M,\tilde g)$ with the image of the isometric
embedding in $(M,g)$.
\begin{enumerate}
\item If $\sigma_0>0$ for $(\tilde M,\tilde g)$, then $(M,g)$ and
  $(\tilde M,\tilde g)$ coincide, and both correspond to the maximal
  globally hyperbolic development of Nariai data. This follows from
  geodesic completeness.
\item If $\sigma_0<0$ for $(\tilde M,\tilde g)$, then $(\tilde
  M,\tilde g)$ is only defined for $t\in I\not=\R$, and $(M,g)$ is an
  extension of $(\tilde M,\tilde g)$. From \Eqref{eq:meancurvatureKS},
  we see that the mean curvature of the $s=const$-surfaces, which are
  Cauchy surfaces of $(\tilde M,\tilde g)$, blows up uniformly exactly
  at the boundary of the embedding. This implies that the solution
  cannot be extended as a globally hyperbolic spacetime.  We conclude
  that $(\tilde M,\tilde g)$ is the maximal globally hyperbolic
  development of Nariai data.  However, since the curvature is
  bounded, it is extendible and the boundary of $(\tilde M,\tilde g)$
  in $(M,g)$ is a Cauchy horizon of topology \SoXSt, generated by
  closed null curves.
\item If $\sigma_0=0$ for $(\tilde M,\tilde g)$, then $(\tilde
  M,\tilde g)$ is defined for all $t\in\R$.  After a change of time
  direction, if necessary, we have that
 \[\lim_{t\rightarrow-\infty}s(t)=-\Phi_0^\prime,\quad
 \lim_{t\rightarrow+\infty}s(t)=+\infty.\] Hence, $(M,g)$ is an
 extension of $(\tilde M,\tilde g)$. In this case, the mean curvature
 of the $s=const$-hypersurfaces stays bounded in the limit
 $t\rightarrow-\infty$. Based on these simple arguments it is not
 possible to conclude that $(\tilde M,\tilde g)$ is the maximal
 globally hyperbolic development of Nariai data with $\sigma_0=0$,
 because we cannot exclude the possibility that there exist other
 extensions which are globally hyperbolic.  We will not discuss this
 issue further here, since, as far as we can see at the moment, the
 case $\sigma_0=0$ is a borderline case of no particular interest for
 us.
\end{enumerate}

Now let us consider the ``generic'' case ${H^{(0)}_*}\not=0$. Then we
can introduce the following new coordinates
\begin{equation}
  \label{eq:defhats}
  \hat s=\frac {\sqrt{G_*}}2 (H^{(0)}_* s+2),\quad 
  \hat\mu=\frac{2}{H^{(0)}_*\sqrt{G_*}}\,\mu.
\end{equation}
In these coordinates, the metric becomes
\[g=2d\hat sd\hat\mu +\underbrace{\frac{(H^{(0)}_*)^2G_*}4 F\left((2\hat
    s/\sqrt{G_*}-2)/H^{(0)}_*\right)}_{=:\hat F(\hat s)} d\hat\mu^2 
+\hat s^2 g_{\St}.\]
After straightforward computations, substituting $F_*$ by means of
the constraint, we find
\begin{subequations}
  \label{eq:KS_SDSForm}
  \begin{equation}  
    \hat F(\hat s)=-1
    +\frac{2S}{\hat s}+\frac{1}3\hat s^2
  \end{equation}
  with
  \begin{equation}
    S:=G_*^{3/2}\left(\frac{1}3
      -\frac 14\dot F_* H^{(0)}_*\right).
  \end{equation}
\end{subequations}
Hence we have shown that the Kantowski-Sachs solutions with
$H^{(0)}_*\not=0$ yield the Schwarzschild-de-Sitter solution with mass
$S$ in Eddington-Finkelstein coordinates.

\begin{figure}[t]
  \centering
  \psfrag{J+}{\scrip}
  \psfrag{J-}{\scrim}
  \psfrag{I}{A}
  \psfrag{II}{B}
  \includegraphics[width=0.8\textwidth]{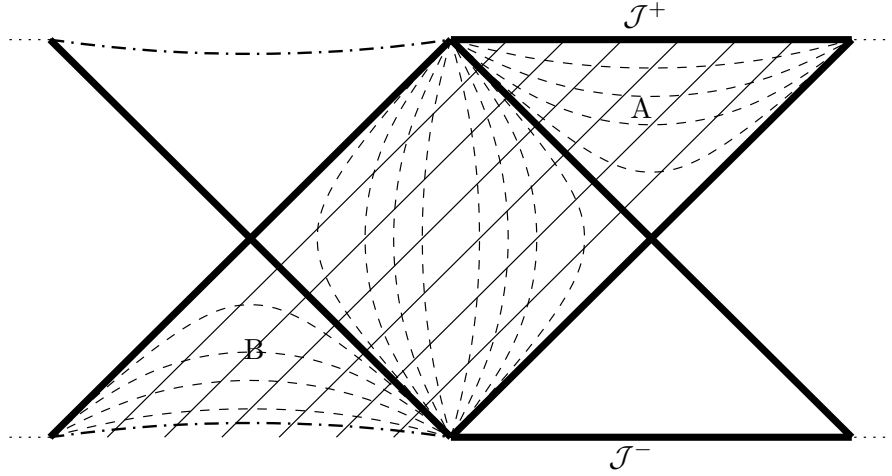}
  \caption{Penrose diagram of Schwarzschild-de-Sitter in the mass
    regime $0<S<1/3$ and $\Lambda=1$}
  \label{fig:SSdSBH}
\end{figure}
Let us make a few comments about the Schwarzschild-de-Sitter solution
by means of \Figref{fig:SSdSBH}. More details are presented in
\cite{beig05}. The figure shows the Penrose diagram of the
Schwarzschild-de-Sitter solution with mass $0<S<1/3$ and $\Lambda=1$
as an example.  Bold diagonal lines denote the event and cosmological
horizons, bold dashed curves refer to the singularity, and bold
horizontal lines mark parts of the conformal boundary $\scrip$ and
$\scrim$. Furthermore, thin dashed curves represent
$s=const$-hypersurfaces and thin non-dashed curves
$\mu=const$-hypersurfaces. The $s=const$-surfaces are spacelike in
regions A and B, but timelike between the cosmological and black hole
event horizons. The cosmological horizon and the event horizon become
Cauchy horizons for the Cauchy development of any
$s=const$-hypersurface in the regions A and B.

We remark that all the results here are consistent with the Birkhoff
theorem for $\Lambda>0$ proved in \cite{Stanciulescu}.

\subsection{Perturbations of Nariai within the Kantowski-Sachs family}
So far we have shown that all Kantowski-Sachs solutions of our initial
value problem with ${H^{(0)}_*}\not=0$ are locally isometric to
subsets of Schwarzschild-de-Sitter solutions with mass $S$ given by
\Eqref{eq:KS_SDSForm}, while for ${H^{(0)}_*}=0$, we get the
generalized Nariai solutions. In order to simplify the discussion
again, we will often go to the universal covers in the following
without further comment.  Let us fix a generalized Nariai solution
$(M,g)$ either by means of the parameters $\Phi_0>0$,
$\Phi_0^\prime\in\R$ or by $F_*>0$, $\dot F_*\in\R$, ${H^{(0)}_*}=0$
related by \Eqref{eq:NariaiIdent}.  Now, leave all these parameters
fixed except for $H^{(0)}_*$, which we give some small but
non-vanishing value. We denote the corresponding Kantowski-Sachs
solution by $(\tilde M,\tilde g)$ and call it the perturbation of
$(M,g)$.

For sufficiently small $|H^{(0)}_*|>0$, we find that $(\tilde M,\tilde
g)$ is a Schwarzschild-de-Sitter solution with mass
\[S=\frac 13+\frac{\sigma_0}8 (H^{(0)}_*)^2
-\Phi_0^\prime\frac{2(\Phi_0^\prime)^2 - 3\sigma_0}{24}(H^{(0)}_*)^3
+O((H^{(0)}_*)^4).\] Note that these and some of the following
formulas have been derived before, cf.\ \cite{Ginsparg83,Dias03}, for
the standard Nariai solution. For sufficiently small $|H^{(0)}_*|>0$,
we have the following conclusions:
\begin{enumerate}
\item Suppose a generalized Nariai solution $(M,g)$ with $\sigma_0>0$
  is given. Then the perturbations $(\tilde M,\tilde g)$ are locally
  isometric to a Schwarzschild-de-Sitter solution with mass $S>1/3$.
\item If $\sigma_0<0$, then $(\tilde M,\tilde g)$ is locally isometric
  a Schwarzschild-de-Sitter solution with $0<S<1/3$.
\item If $\sigma_0=0$, then $(\tilde M,\tilde g)$ has the following
  properties:
  \begin{enumerate}
  \item If $\Phi_0^\prime H^{(0)}_*>0$: locally isometric to a
    Schwarzschild-de-Sitter solution with $0<S<1/3$.
  \item If $\Phi_0^\prime H^{(0)}_*<0$: locally isometric to a
    Schwarzschild-de-Sitter solution with $S>1/3$.
  \item The case $\Phi_0^\prime H^{(0)}_*=0$ can be excluded.
  \end{enumerate}
\end{enumerate}
It is an interesting result that the three isometry classes of
generalized Nariai solutions yield quite different perturbations
despite of the fact that they are locally isometric.

For $(\tilde M,\tilde g)$ for given finite $s$ and $\mu$ and
sufficiently small $|H^{(0)}_*|>0$, we find
\begin{align*}
  \hat s&=1+\frac{\Phi_0^\prime + s}2 H^{(0)}_*+O((H^{(0)}_*)^2),\\
  \hat\mu&=\mu\left(\frac{2}{H^{(0)}_*}-\Phi_0^\prime
    -\frac{2 (\Phi_0^\prime) +\sigma_0}4 H^{(0)}_*+O((H^{(0)}_*)^2)\right).
\end{align*}
Note that this suggests that generalized Nariai solutions can be
considered as singular limits of Schwarzschild-de-Sitter solutions
with masses as discussed above. That is, in the limit
$H^{(0)}_*\rightarrow 0$, we have
\[M\rightarrow 1/3,\quad \hat s\rightarrow 1,\]
and further, for $H^{(0)}_*\searrow 0$, one gets
\[\hat\mu\rightarrow
\begin{cases}
  -\infty & \mu<0\\
  0       & \mu=0\\
  \infty  & \mu>0.
\end{cases}
\]
For $H^{(0)}_*\nearrow 0$, the signs for the limits of $\hat\mu$ turn
around.  We will not consider the consequences of this, but see for
instance \cite{Ginsparg83,gibbons77}. However, note that $\mu$ and $s$
have to be finite in order to give sense to these expansions, and thus
they do not yield information about the asymptotics of the generalized
Nariai solution. This and the fact that the limit of $\hat\mu$ is
singular in this ``Nariai limit of the Schwarzschild-de-Sitter class''
is obvious from the formulas above, but often not pointed out clearly
in the literature. For instance, the author of \cite{Bousso03} is led
to the statement that the Nariai solution is ``the largest possible
black hole in de Sitter space'', i.e.\ a Schwarzschild-de-Sitter
solution with mass $M=1/3$, by the analogous considerations in
\cite{Ginsparg83}. At least from our point of view, this statement is
misleading.

In order to compare the global properties of the perturbation $(\tilde
M,\tilde g)$ with those of the unperturbed Nariai solution $(M,g)$,
let us restrict to the example case $\sigma_0<0$. This is not the
standard case in the literature, but will play a particular role in
the following paper \cite{beyer09:Nariai2}. Let us consider
\Figref{fig:SSdSBH} for the Penrose diagram of the
Schwarzschild-de-Sitter solution with $0<S<1/3$. We can change the
time direction so that $\Phi_0^\prime>0$ (note that $\Phi_0^\prime=0$
is not allowed in this class) for the unperturbed Nariai solution
$(M,g)$. Now let $H^{(0)}_*$ be sufficiently small and positive. In
this case, the initial hypersurface of our Kantowski-Sachs initial
value problem given by $s=0$ corresponds to $\hat s=\hat s_0>1$. From
the results derived above and the well-known root structure of $\hat
F$, it follows that the initial hypersurface must correspond to a
$\hat s=const$-Cauchy surface of region $A$ in
\Figref{fig:SSdSBH}. Hence, we can say the following in this case:
\begin{itemize}
\item In the future time direction, while the unperturbed Nariai
  solution $(M,g)$ has Nariai asymptotics, the perturbation $(\tilde
  M,\tilde g)$ has a smooth future conformal boundary \scrip. In other
  words, while for the original solution the volume of the \St-factor
  is constant in time, an arbitrarily small initial positive expansion
  of the \St-factor determined by a small $H^{(0)}_*>0$ leads to
  accelerated expansion of the \St-factor.  Together with the
  accelerated expansion of the \So-factor present also in the
  unperturbed solution, this yields the existence of a smooth future
  conformal boundary \scrip.
\item In the past time direction, both $(M,g)$ and the perturbation
  $(\tilde M,\tilde g)$ develop a Cauchy horizon, which in the case of
  the perturbation corresponds to the cosmological horizon. Thus the
  maximal globally hyperbolic extension of the data on the
  $s=0$-hypersurface is extendible.
\end{itemize}
The case $H^{(0)}_*<0$ can be discussed similarly. The relevant region
in \Figref{fig:SSdSBH} is now marked as region B, but note that we
have to go backwards with respect to the Schwarzschild-de-Sitter time
due to the definition of $\hat s$ in \Eqref{eq:defhats}. For the
perturbation $(\tilde M,\tilde g)$ the initial expansion of the
\St-factor is now negative and the result is that the future Nariai
asymptotics of $(M,g)$ turn into a curvature singularity for the
perturbation $(\tilde M,\tilde g)$. The singularity is of cigar type
according to the classification in \cite{Wainwright}, because,
although the volume of the spatial surfaces shrinks to zero, the
volume of the \So-factor becomes infinite at the singularity. In the
past, both $(M,g)$ and $(\tilde M,\tilde g)$ develop a Cauchy
horizon. By means of similar arguments, the cases $\sigma_0\ge 0$ can
also be studied.

%%% Local Variables: 
%%% mode: latex
%%% TeX-master: "paper"
%%% End: 

\section{Summary and outlook}
\label{sec:summary}
This paper is devoted to the discussion of the cosmic no-hair
conjecture and its relation to the Nariai solutions. Our discussion
suggests that Nariai asymptotics is a non-generic phenomenon for
solutions of Einstein's field equations in vacuum with a positive
cosmological constant. We have described the relation of cosmic
no-hair and conformal boundaries first, and then analyzed the global
properties of Nariai solutions in order to prove the non-existence of
smooth conformal boundaries. We point out that the analysis of
conformal boundaries serves as a good general diagnostic tool in order
to find out about the cosmic no-hair picture, as we also see in
\cite{beyer09:Nariai2}.  Then, we have explained the instability of
the Nariai solutions in the spatially homogeneous case in a precise
manner.

Beyond the problem of cosmic no-hair, the particular realization of
the instability of the Nariai solution in terms of the sign of the
quantity $H_*^{(0)}$ makes it tempting to study the following question
in the spatially inhomogeneous case. Can we exploit the instability of
the Nariai solutions, which we understand in the spatially homogeneous
case, and construct arbitrarily complicated cosmological black hole
solutions by making $H_*^{(0)}$ spatially dependent on the initial
hypersurface? In principle, we are interested in studying generic
inhomogeneous perturbations of the Nariai solutions by prescribing a
function $H_*^{(0)}$ without any symmetries. However, this seems
hopeless in practice. A systematic approach would be to reduce the
symmetry step by step. The first systematic step for the study of
inhomogeneous perturbations is the spherically symmetric case; i.e.\
we give up the homogeneity along the $\So$-factor of the spatial
manifold, but keep all the symmetries of the $\St$-factor. Much is
known about this case locally, for instance by the Birkhoff theorem in
\cite{Stanciulescu}. But to our knowledge, there are no rigorous
results of a global nature which are relevant for our questions
here. Nevertheless, Bousso in \cite{Bousso03} claims that cosmological
black holes can be constructed as above in the spherically symmetric
case.

We have decided not to proceed with the spherically symmetric case in
our second paper \cite{beyer09:Nariai2}, but rather with Gowdy
symmetry. This symmetry is characterized by the presence of two
spatial Killing vector fields. It has turned out that it is very
difficult to handle this class analytically, and to our knowledge, no
cosmological black hole solutions in this class have been found
before.  This is one motivation for us to proceed with this class by
means of numerical techniques.  We have developed numerical techniques
which can also be applied to the \SoXSt-Gowdy class of solutions in
\cite{beyer08:code}. Our approach and the first steps in the
investigation of the instability of Nariai asymptotics and the
construction of cosmological black hole solutions within the Gowdy
class are presented in \cite{beyer09:Nariai2}.

%%% Local Variables: 
%%% mode: latex
%%% TeX-master: "paper"
%%% End: 

\section{Acknowledgments}
This work was supported in part by the Göran Gustafsson Foundation,
and in part by the Agence Nationale de la Recherche (ANR) through the
Grant 06-2-134423 entitled Mathematical Methods in General Relativity
(MATH-GR) at the Laboratoire J.-L. Lions (Universit\'e Pierre et Marie
Curie).  Some of the work was done during the program ``Geometry,
Analysis, and General Relativity'' at the Mittag-Leffler institute in
Stockholm in fall 2008. I would like to thank in particular Helmut
Friedrich and Hans Ringström for helpful discussions and explanations.

%%% Local Variables: 
%%% mode: latex
%%% TeX-master: "paper"
%%% End:

\bibliography{bibliography}
\end{document}